\documentclass[aps,prd,10pt,twocolumn,nofootinbib,groupedaddress,showpacs]{revtex4-1}
\usepackage[latin1]{inputenc}
\usepackage{placeins}
\usepackage{booktabs}
\usepackage{dcolumn}
\usepackage{array}
\usepackage{longtable}
\usepackage{amsmath,amsfonts,amsthm,amssymb}
\usepackage{bm,graphicx,graphics,color}
\usepackage{epsf,epsfig}
\usepackage[all]{xy}
\usepackage{amsmath}
\usepackage{amsfonts}
\usepackage{amssymb}
\usepackage{graphicx}%
\providecommand{\U}[1]{\protect\rule{.1in}{.1in}}

\begin{document}
\title{Weyl frames and canonical transformations in geometrical scalar-tensor
theories of gravity }

\author{A. B. Barreto}
\email{adriano.barreti@caxias.ifrs.edu.br}
\affiliation{Instituto Federal de Educação, Ciência e Tecnologia do Rio Grande do Sul,
95043-700, Caxias do Sul-RS, Brazil \linebreak} 
\affiliation{Departamento de F\'{i}sica,Universidade Federal da Para\'{i}ba, C. Postal 5008, 58051-970, Jo\~ao Pessoa-PB, Brazil}

\author{M. L. Pucheu}
\email{mlaurapucheu@fisica.ufpb.br}
\affiliation{Departamento de F\'{i}sica,Universidade Federal da Para\'{i}ba, C. Postal 5008, 58051-970, Jo\~ao Pessoa-PB, Brazil}

\author{C. Romero}
\email{cromero@fisica.ufpb.br}
\affiliation{Departamento de F\'{i}sica,Universidade Federal da Para\'{i}ba, C. Postal 5008, 58051-970, Jo\~ao Pessoa-PB, Brazil}

\begin{abstract}
We consider scalar-tensor theories of gravity defined in Weyl integrable
space-time and show that in the ADM formalism Weyl transformations
corresponding to change of frames induce canonical transformations between
different representations of the phase space. In this context, we discuss the
physical equivalence of two distinct Weyl frames at the classical level.

\end{abstract}
\maketitle

\date{\today}


\section{Introduction}

As is well known, the Hamiltonian formalism has proved to be a powerful tool
in the study of the dynamics of classical systems. Its application to general
relativity, as well as to other theories of gravity, based on the so-called
ADM formalism \cite{ADM}, \ has set up the basis for several approaches to a
quantum theory of gravity \cite{QG}. Certainly, a prominent aspect of the
Hamiltonian formalism is related to canonical transformations, the very
special class of transformations defined in the phase space which preserves
the form of Hamilton equations. Clearly, dynamical systems leading to Hamilton
equations which may be related by canonical transformations are to be regarded
as being physically equivalent \cite{goldstein}.

In a somewhat different context, namely, that of scalar-tensor theories of
gravity, the issue of physical equivalence appears when different frames are
used to write the field equations. Usually these frames are related by a set
of transformations involving the metric and the scalar field. In the case of
Brans-Dicke gravity, two frames, in particular, are considered as important
for the mathematical formulation of the theory: the Einstein and Jordan frames
\cite{BD}. The question of which should be regarded as the physical frame is
still a matter of debate \cite{Faraoni}.

It turns out that, with respect to scalar-tensor theories, the original
approach assumes, as in general relativity, that the space-time manifold is
Riemannian. On the other hand, it has been shown recently that when the
Palatini variational method is applied to derive the field equations from the
action, then in a wide class of scalar-tensor theories, a non-Riemannian
compatibility condition between metric and affine connection appears quite
naturally \cite{GBD} (For a more general result, see\ \cite{Burton1997}). In a
certain sense, this condition seems to establish the space-time geometry from
first principles, the space-time manifold being dynamically determined by the
particular coupling of the scalar field in the gravitational sector. In the
case of Brans-Dicke theory, the mentioned procedure leads to what has been
called a \textit{Weyl integrable space-time}, a particular version of the
geometry conceived by H. Weyl in his attempt to unify gravity and
electromagnetism \cite{Weyl}\ 

Now, Weyl geometry is one of the simplest generalizations of Riemann geometry,
in which the metric compatibility condition is weakened. This was the way
Weyl devised to introduce a covariant vector field $\sigma_{\mu}$, which bears
amazing similarity with the electromagnetic 4-potential. Weyl also introduced
the tensor $F_{\mu\nu}=\partial_{\mu}\sigma_{\nu}-\partial_{\nu}\sigma_{\mu}$,
which he interpreted as representing a kind of \textit{length curvature}. As a
consequence of the modification in the Riemannian compatibility condition, the
covariant derivative of the metric tensor does not vanish, as in Riemannian
geometry, and the length of vectors parallel transported along a curve may
change. Weyl's compatibility condition is given by $\nabla_{\alpha}g_{\mu\nu
}=\sigma_{\alpha}g_{\mu\nu}$, and is invariant under the conformal
transformation $g_{\mu\nu}\rightarrow\overline{g}_{\mu\nu}=e^{f}g_{\mu\nu}$
and $\sigma_{\mu}\rightarrow\overline{\sigma}_{\mu}=\sigma_{\mu}+\partial
_{\mu}f$, where $f$ is an arbitrary scalar function \cite{Weyl geometry}.
\ These findings are considered by some authors as the "dawning" of\ modern
gauge theories \cite{afriat}. If $F_{\mu\nu}=0$ (null second curvature) then
there is no electromagnetic field. In this case, there exists a scalar field
$\phi$, such that $\sigma_{\mu}=\partial_{\mu}\phi$, and, instead of a vector
field, we are left with a scalar field $\phi$, which, in addition to the
metric, is the fundamental object that characterizes this geometry. A
space-time endowed with this particular version of Weyl geometry is known
as\textit{ Weyl Integrable Space-Time }\cite{WIST}.

In this article, we shall show that, when we consider the ADM formalism for
scalar-tensor theories, then Weyl transformations induce canonical
transformations between different representations of the phase space. We
obtain a generating function corresponding to the canonical transformations,
thereby showing the physical equivalence of two distinct Weyl frames at the
classical level. We revisit, in this way, the discussion about the physical
equivalence between Jordan and Einstein frames, now from the point of view of
Weyl geometry, in which affine geodesics and the concept of proper time are
invariant under frame transformations.

\section{Hamiltonian formalism and Weyl frames }

Let us consider the gravitational sector of Brans-Dicke action
\begin{equation}
\mathcal{S}=\int d^{4}x\sqrt{-g}e^{-\phi}\left(  R+\omega\phi_{,\mu}\phi
^{,\mu}\right)  , \label{S_WIST}%
\end{equation}
where $R$ is the Ricci scalar calculated from the Weyl connection, and
$\omega$ denotes a free dimensionless parameter.

As already mentioned, Weyl integrable geometry is an extension of Riemannian
geometry. In the case of Weyl integrable space-time, the transformations
mentioned above reduce to \footnote{In the literature, these transformations
are also referred to either as \textit{gauge transformations }or \textit{
change of frames}.}%

\begin{align}
\bar{g}_{\mu\nu}  &  =e^{f}g_{\mu\nu},\label{weyltransf}\\
\bar{\phi}  &  =\phi+f,\nonumber
\end{align}

Let us now restrict ourselves to homogeneous and isotropic cosmological
models, with the Friedman-Lemaître-Robertson-Walker line element given by
\begin{equation}
ds^{2}=N^{2}(t)dt^{2}-a^{2}(t)\left[  \frac{dr^{2}}{1-kr^{2}}+r^{2}\left(
d\theta^{2}+\sin^{2}\theta d\varphi^{2}\right)  \right]  ,\label{dsFW}%
\end{equation}
where $N(t)$ denotes the lapse function and $a(t)$ is the cosmic scale factor.
It is not difficult to see that, after neglecting surface terms, the reduced
action of the Lagrangian corresponding to (\ref{dsFW}) is
\begin{equation}
L=e^{-\phi}\left[  \left(  \omega-\frac{3}{2}\right)  \frac{a^{3}}{N}\dot
{\phi}^{2}+6\left(  kNa-\frac{a}{N}\dot{a}^{2}+\frac{a^{2}}{N}\dot{a}\dot
{\phi}\right)  \right]  .\label{LagranFW}%
\end{equation}
On the other hand, the canonical momenta will be given by
\begin{align}
p_{a} &  =\frac{6e^{-\phi}}{N}\left(  a^{2}\dot{\phi}-2a\dot{a}\right)
,\label{Pa}\\
p_{\phi} &  =\frac{e^{-\phi}}{N}\left[  \left(  2\omega-3\right)  a^{3}%
\dot{\phi}+6a^{2}\dot{a}\right]  .\label{Pphi}%
\end{align}
Thus the total Hamiltonian of the model can be written as
\begin{equation}
H=N\mathcal{H},\label{totalhamiltonian}%
\end{equation}
with the super-Hamiltonian constraint being
\begin{equation}
\mathcal{H}=\frac{e^{\phi}}{4\omega a}\left[  \frac{(3-2\omega)}{12}{p_{a}%
}^{2}+\frac{{{p_{\phi}}^{2}}}{a^{2}}+\frac{p_{a}p_{\phi}}{a}\right]
-6kae^{-\phi}.\label{hamiltonian}%
\end{equation}

At this point, let us transform the action (\ref{S_WIST}) (written in the Weyl
frame $(g,\phi)$) by performing the general Weyl transformations
(\ref{weyltransf}). It is not difficult to verify that (\ref{S_WIST}), turns
into the new action
\begin{equation}
\mathcal{\bar{S}}=\int d^{4}x\sqrt{-\bar{g}}e^{-\bar{\phi}}\left(  \bar
{R}+\omega\bar{\phi}_{,\mu}\bar{\phi}^{,\mu}-2\omega\bar{\phi}_{,\mu}f^{,\mu
}+\omega f_{,\mu}f^{,\mu}\right)  ,\label{action transformed}%
\end{equation}
now written in the new frame $(\bar{g},\bar{\phi})$. We now rewrite the FLRW
line element as
\begin{equation}
d\bar{s}^{2}=\bar{N}^{2}dt^{2}-\bar{a}^{2}(t)\left[  \frac{dr^{2}}{1-kr^{2}%
}+r^{2}\left(  d\theta^{2}+\sin^{2}\theta d\varphi^{2}\right)  \right]  .
\end{equation}
In the context of the Hamiltonian formalism it seems reasonable to regard $f$
as a function of $\bar{a}$ and $\bar{\phi}$, i. e., $f\equiv f(\bar{a}%
,\bar{\phi})$. It is easy to see that this assumption leads to the reduced
Lagrangian
\begin{align}
\bar{L} &  =\frac{e^{-\bar{\phi}}}{\bar{N}}\bigg\{\bar{a}\left(  \omega\bar
{a}^{2}{f_{,\bar{a}}}^{2}-6\right)  \dot{\bar{a}}^{2}+2\bar{a}^{2}\left[
3-\omega\bar{a}\left(  1-f_{,\bar{\phi}}\right)  f_{,\bar{a}}\right]
\times\nonumber\\
&  \times\dot{\bar{a}}\dot{\bar{\phi}}+\bar{a}^{3}\left[  \omega\left(
1-f_{,\bar{\phi}}\right)  ^{2}-\frac{3}{2}\right]  \dot{\bar{\phi}}^{2}%
+6k\bar{a}\bar{N}^{2}\bigg\},\nonumber\\
&
\end{align}
where we have defined $f_{,\bar{a}}=\frac{\partial f}{\partial\bar{a}}$ and
$f_{,\bar{\phi}}=\frac{\partial f}{\partial\bar{\phi}}$. The canonical momenta
in the transformed frame are given by
\begin{equation}
p_{\bar{a}}=\frac{2\bar{a}e^{-\bar{\phi}}}{\bar{N}}\left\{  \dot{\bar{a}%
}\left(  \omega\bar{a}^{2}{f_{,\bar{a}}}^{2}-6\right)  +\bar{a}\dot{\bar{\phi
}}\left[  3-\omega\bar{a}\left(  1-f_{,\bar{\phi}}\right)  f_{,\bar{a}%
}\right]  \right\}  ,
\end{equation}%
\begin{equation}
\begin{aligned} p_{\bar{\phi}} = \frac{2 \bar{a}^2 e^{-\bar{\phi}}}{\bar{N}} \bigg\lbrace \dot{\bar{a}} \left[ 3- \omega \bar{a}\left(1-f_{,\bar{\phi}} \right) f_{,\bar{a}} \right] + \\ \bar{a} \dot{\bar{\phi}} \left[\omega \left(1- f_{,\bar{\phi}} \right)^2 - \frac{3}{2}\right] \bigg\rbrace. \end{aligned}
\end{equation}
while the super-Hamiltonian constraint expressed in the new variables takes
the form
\begin{equation}
\begin{aligned} \bar{\mathcal{H}}= \frac{e^{\bar{\phi}}}{12 \omega \bar{a} \left[\bar{a} f_{,\bar{a}}+ 2 (f_{,\bar{\phi}}-1) \right]} \biggl\{ \left[ 3- 2\omega (f_{,\bar{\phi}}-1)^2\right] {p_{\bar{a}}}^2 + \\ + 2 \left(6- \omega \bar{a}^2 {f_{\bar{a}}}^2 \right) \frac{{p_{\bar{\phi}}}^2}{\bar{a}^2} + 4 \left[3+ \omega \bar{a} f_{,\bar{a}}\left(f_{,\bar{\phi}}-1 \right) \right] \frac{p_{\bar{a}}p_{\bar{\phi}}}{\bar{a}} \biggr\} +\\ - 6 k \bar{a} e^{-\bar{\phi}}. \end{aligned}\label{transfHamiltonian}%
\end{equation}

Since $\bar{\mathcal{H}}$ and $\mathcal{H}$ come from two distinct Lagrangians
related by the Weyl transformation (\ref{weyltransf}), it seems interesting to
investigate whether there exists a transformation in the phase space that
links them. According with the Hamiltonian mechanics, such a transformation,
if it exists at all, must be a canonical transformation. In the next section,
we shall show that, indeed, Weyl transformations induce canonical
transformations which relate the generalized coordinates in both Weyl frames.

\subsection{Change of frames as canonical transformations}

In what follows, we shall look for a change of frames that preserves the form
of Hamilton equations. In other words, we look for a transformation that links
the canonical coordinates and takes the total Hamiltonian $H=N\mathcal{H}$
into $\bar{H}=\bar{N}\bar{\mathcal{H}}$. This can easily be done by computing the
relevant Poisson brackets and showing that they are preserved. Let us first
consider the following class of transformations
\begin{align}
\bar{a} &  =e^{\frac{f}{2}}a,\nonumber\label{CanTransf}\\
\bar{\phi} &  =\phi+f,\nonumber\\
p_{\bar{a}} &  =\frac{2e^{-\frac{f}{2}}}{2+af_{,a}}\left(  p_{a}-f_{,a}%
p_{\phi}\right)  ,\nonumber\\
p_{\bar{\phi}} &  =\frac{1}{1+f_{,\phi}}\left(  p_{\phi}-\frac{af_{,\phi}}%
{2}p_{a}\right)  .
\end{align}
It is not difficult to check that these transformations are canonical in two
cases: i) $f_{,a}=0$ and $f(\phi)\neq-\phi$, and ii) $f_{\phi}\equiv
\frac{\partial f}{\partial\phi}=0$ and $f(a)\neq\ln{a^{2}}$. These relate, in
principle, two distinct gravitational theories whose actions are defined in
two distinct Weylian frames connected by (\ref{weyltransf})\footnote{Although
the lapse function $N$ plays the role of a Lagrange multiplier in the
Hamiltonian formalism, its redefinition is also required for a full
identification between the Hamiltonians by means of (\ref{CanTransf}). Thus we
have set $N=e^{-f/2}\bar{N}$.}. On the other hand, because\ these
transformations, which are canonical, originated from a change in the Weyl
frames, is sufficient to guarantee the physical equivalence between the
two\ theories, at least at the classical level. Unfortunately this equivalence
between frames cannot be taken further. Indeed, at the quantum level, to say
the least, it is still unclear whether classically equivalent systems (related
by canonical transformations) lead to quantum equivalent systems
\cite{Anderson}.

\subsection{Generating functions}

Another way to show that\ the transformations (\ref{CanTransf}) are canonical
is to obtain explicitly its generating function. For this purpose, let us
consider the first case examined previously, that is, when $f\equiv f(\phi)$,
with $f\neq-\phi$. The Weyl canonical transformations are given by%

\begin{align}
\bar{a}  &  =e^{\frac{f(\phi)}{2}}a,\nonumber\label{CanTransffphi}\\
\bar{\phi}  &  =\phi+f(\phi),\nonumber\\
p_{\bar{a}}  &  =e^{-\frac{f(\phi)}{2}}p_{a},\nonumber\\
p_{\bar{\phi}}  &  =\frac{1}{1+\frac{df(\phi)}{d\phi}}\left(  p_{\phi}%
-\frac{a\frac{df(\phi)}{d\phi}}{2}p_{a}\right)  ,
\end{align}
and it is straightforward to verify that the generating function of this
transformation is
\begin{equation}
G_{1}=f(\phi)\left(  p_{\phi}+\frac{ap_{a}}{2}\right)  .
\end{equation}

Let us now consider the second case, in which $f_{\phi}=0$. For simplicity,
let us choose as a particular example, $f\equiv f(a)=e^{a}$. The above
transformations then reduce to
\begin{align}
\bar{a}  &  =\exp{\left(  e^{a}/2\right)  }a,\label{CanTransffa}\\
\bar{\phi}  &  =\phi+e^{a},\nonumber\\
p_{\bar{a}}  &  =\frac{2\exp{\left(  -e^{a}/2\right)  }}{2+ae^{a}}\left(
p_{a}-e^{a}p_{\phi}\right)  ,\nonumber\\
p_{\bar{\phi}}  &  =p_{\phi},
\end{align}
and are generated by the function
\begin{equation}
G_{2}=e^{a}\left(  p_{\phi}+\frac{ap_{a}}{2}\right)  .
\end{equation}

\subsection{A particular example: the Riemann frame}

Let us next consider the particular Weyl transformation that leads to the
frame in which the transformed geometrical scalar field $\widetilde{\phi}%
\ $vanishes, i.e., the so-called Riemann frame, in which the Riemannian
compatibility condition
\begin{equation}
\nabla_{\alpha}\widetilde{g}_{\mu\nu}=0
\end{equation}
is recovered (For details, see \cite{GBD}). Clearly, this is carried out
simply by taking $f=-\phi$ $\ $in (\ref{weyltransf}). In this case, it is easy
to see that the transformed action takes the form
\begin{equation}
\mathcal{\tilde{S}}=\int d^{4}x\sqrt{-\tilde{g}}\left(  \tilde{R}+\omega
\tilde{g}^{\mu\nu}\phi_{,\mu}\phi_{,\nu}\right)  ,\label{RF-Action}%
\end{equation}
where $\tilde{g}^{\mu\nu}=e^{\phi}g^{\mu\nu}$, and the Ricci scalar $\tilde
{R}$ is defined purely in terms of the metric $\widetilde{g}_{\mu\nu}$
\footnote{It is important to note that, in the action written in the Riemann
frame, $\phi$ is no longer the Weyl geometrical field, and its appearance here
is due simply to the particular choice of the function $f$ \ in the Weyl
transformation, namely, $f=-\phi.$}. It is worth noting that when
$\omega=\frac{1}{2}$ (\ref{RF-Action}) is formally equivalent to the
Hilbert-Einstein action with a massless minimally coupled scalar field
\cite{GBD}.

Again, from the line element%
\begin{equation}
d\tilde{s}^{2}=\tilde{N}^{2}(t)dt^{2}-\tilde{a}^{2}(t)\left[  \frac{dr^{2}%
}{1-kr^{2}}+r^{2}\left(  d\theta^{2}+\sin^{2}\theta d\varphi^{2}\right)
\right]  , \label{dsFR}%
\end{equation}
the reduced Lagrangian is easily seen to be given by
\begin{equation}
\tilde{L}=6\left(  k\tilde{a}\tilde{N}-\frac{\tilde{a}}{\tilde{N}}\dot
{\tilde{a}}^{2}\right)  +\omega\frac{\tilde{a}^{3}}{\tilde{N}^{2}}\dot
{\tilde{\phi}}^{2}. \label{LagrFR}%
\end{equation}
A brief comment about (\ref{LagrFR}) is in order. As $\tilde{L}$ does not
depend explicitly on $\tilde{\phi}$, it follows that its conjugate canonical
momentum $p_{\tilde{\phi}}$ is conserved. On the other hand, the canonical
momenta are
\begin{align}
p_{\tilde{a}}  &  =-\frac{12}{\tilde{N}}\tilde{a}\dot{\tilde{a}},\label{Pb}\\
p_{\tilde{\phi}}  &  =\frac{2}{\tilde{N}}\omega\tilde{a}^{3}\dot{\tilde{\phi}%
}, \label{Psigma}%
\end{align}
while the super-Hamiltonian reads
\begin{equation}
\mathcal{\tilde{H}}=-\frac{{p_{\tilde{a}}}^{2}}{24\tilde{a}}+\frac
{{p_{\tilde{\phi}}}^{2}}{4\omega\tilde{a}^{3}}-6k\tilde{a}. \label{SHRiemann}%
\end{equation}

Finally, the canonical transformations in the phase space that relate the
canonical variables of the two actions (\ref{S_WIST}) and (\ref{RF-Action})
are
\begin{align}
\tilde{a}  &  =ae^{-\phi/2},\nonumber\label{TransfCanRiemann}\\
\tilde{\phi}  &  =\phi,\nonumber\\
p_{\tilde{a}}  &  =p_{a}e^{\phi/2},\nonumber\\
p_{\tilde{\phi}}  &  =p_{\phi}+\frac{a}{2}p_{a},
\end{align}
with the new lapse function being given by $\tilde{N}=Ne^{-\phi/2}$. The
generating function in this case is simply
\begin{equation}
\tilde{G}=\frac{1}{2}\tilde{\phi}\tilde{a}p_{\tilde{a}}. \label{FGC}%
\end{equation}

\section{Final remarks\label{sec6}}

In this work we have investigated the problem of how Weyl transformations
behave when viewed at the level of the Hamiltonian formulation of gravity in
the case of scalar-tensor theories. Starting from the original action of
Brans-Dicke theory, in which the underlying space-time is assumed to have the
geometric structure of Weyl integral space-time, we carry out a class of Weyl
frame transformations which induce changes in the reduced Hamiltonian of the
original action. We then obtain the unexpected result that the two
Hamiltonians are related by a canonical transformation. The physical
equivalence between the actions, particularly when the Weyl transformation
leads to the Riemann frame is, in a certain way, consistent with the recent
interpretation of what physical equivalence means when geometrical
scalar-tensor theories are viewed in different frames \cite{GBD}. By no means
can this equivalence be extended to the quantum level \cite{Anderson}. This is
the case, for instance, when we are working out  the canonical quantization of
classical cosmological models in the framework of quantum cosmology. As far as
we know, \ the physical equivalence between frames at the quantum level is
still an open question.
\section*{Acknowledgments}
The authors thank CAPES and CNPq for financial support.

\end{document}